\documentclass[12pt]{article}
\usepackage{latexsym,amssymb,amsmath, bm}
\usepackage{graphicx} 
\usepackage{hyperref}
\usepackage[sort&compress]{natbib}

\def \P {\mathrm{P}}
\def \Re {\mathbb{R}}
\def \ra {\rightarrow}
\def \oo {\infty}
\def \Chat{{\widehat C}}
\def \Cbar {{\overline C}}
\def \uu {\bm{u}} 
\def \T {\mathrm{T}}
\def \tailo {\kappa}
\def \a {\alpha}
\def \RV {\mbox{\rm{RV}}}
\def \al {\alpha} 
\def \xx {\bm{x}}
\def \one {\bm{1}} 
\def \wbf {\bm{w}}
\def \tailorder {\kappa}
\def \lm {\lambda}
\def \XX {\bm{X}}
\def \eqd {\buildrel d \over =}
\def \zz {\bm{z}}
\def \lmm {\left[}
\def \rmm {\right]}
\def \Fbar {{\overline F}}
\def \MDA {\mbox{\rm{MDA}}}
\def \pf {\noindent \emph{Proof: }}
\def \QED {\hfill $\square$}
\def \Gbar {{\overline G}}
\def \E {\mathrm{E}}
\def \G {\Gamma}

\def \FRE {\mbox{\rm{Fr\'{e}chet}}}
\def \lss {\left(}
\def \rss {\right)}
\def \lbb {\left\{}
\def \rbb {\right\}}
\def \half {{\textstyle \frac{1}{2}}}
\def \be {\beta}

\newtheorem{definition}{\bf Definition}

\newtheorem{example}{\bf Example}
\newtheorem{exm}[example]{\bf Example}

\newtheorem{theorem}{\bf Theorem}
\newtheorem{thm}[theorem]{\bf Theorem}
\newtheorem{prop}[theorem]{\bf Proposition}


\begin{document}

\title{Intermediate Tail Dependence: A Review and Some New Results}

\author{Lei Hua\footnote{Corresponding author: {\small\texttt{hua@math.niu.edu}}, Division of Statistics, Northern Illinois University, DeKalb, IL, 60115,
United States.}
\and
Harry Joe\footnote{{\small\texttt{Harry.Joe@ubc.ca}}, Department of Statistics, University of British Columbia, Vancouver, BC, V6T1Z4,
Canada.}
}
\date{}
\maketitle

\begin{abstract}
The concept of intermediate tail dependence is useful if one wants to
quantify the degree of positive dependence in the tails when there is
no strong evidence of presence of the usual tail dependence. We first
review existing studies on intermediate tail dependence, and then we
report new results to supplement the review. Intermediate tail dependence
for elliptical, extreme value and Archimedean copulas are reviewed
and further studied, respectively. For Archimedean copulas, we not only
consider the frailty model but also the recently studied scale mixture
model; for the latter, conditions leading to upper intermediate tail
dependence are presented, and it provides a useful way to simulate
copulas with desirable intermediate tail dependence structures.


\end{abstract}


\section{Introduction}\label{sec-intro}

For applications in many areas such as environmetrics, actuarial science and quantitative finance, in addition to prudent examinations of univariate margins, careful modeling of various dependence patterns in corresponding distributional tails is often very important. For statistical modeling of dependence structures between random variables, a very useful approach is to employ a copula function to combine the univariate margins together to get the joint distribution. A copula $C:[0,1]^d \ra [0,1]$ for a $d$-dimensional random vector can be defined as $C(u_1, \dots, u_d)= F(F_1^{-1}(u_1), \dots, F_d^{-1}(u_d))$, where $F$ is the joint cumulative distribution function (cdf), $F_i$ is the univariate cdf for the $i$th margin, and $F^{-1}_i$ is the generalized inverse function defined as $F^{-1}_i(u) = \inf\{x: F_i(x) \geq u\}$. To avoid technical complexity, throughout the article, the univariate cdfs $F_i$'s are assumed to be supported on $[0,\oo)$, and $F_i$'s are continuous and thus the copula $C$ is unique due to Sklar's theorem \citep{Sklar1959}. We refer the readers to \cite{Joe1997} and \cite{Nelsen2006} for references of copulas. Moreover, all distribution functions and density functions are assumed to be ultimately monotone to the left and right endpoints; this condition is very mild and satisfied by all the commonly used distributions.

The so-called tail dependence parameters (also called tail dependence coefficients or tail dependence index) have been studied as a summary quantity to capture the degree of tail dependence. Let $\lambda$ be the upper tail dependence parameter, then $\lambda := \lim_{u \ra 0^+}\Chat(u,\dots,u)/u$ (provided that the limit exists), where $\Chat$ is the survival copula of $C$; that is, $\Chat(u_1,\dots,u_d): = \Cbar(1-u_1,\dots, 1-u_d)$, and $\Cbar$ is the survival function of $C$ and defined as $\Cbar(u_1, \dots, u_d):= 1+\sum_{\emptyset \neq I \subseteq \{1,\dots,d\}}(-1)^{|I|}C_I(u_i, i \in I)$, where $C_I$ is the copula for the $I$-margin. 

The concept of intermediate tail dependence arises when one wants to quantify the strength of such dependence in the tails but the usual tail dependence parameter $\lambda=0$. In this case, one needs to find another quantity to capture the strength of dependence in the tails. 

In \cite{Hua2011b}, the concept of tail order ($\kappa$) is suggested as a quantity to capture the leading information of dependence in the tails when $\lambda = 0$. When we look at the decay of a copula function along the diagonal as a function in $u$, then a mild assumption is that $C(u,\dots,u) \sim u^\kappa \ell(u)$ as $u \ra 0^+$, where $\ell$ is a slowly varying function \citep{Bingham1987} and the notation $g \sim h$ means that the functions $g$ and $h$ are asymptotically equivalent; that is, $\lim_{t \ra t_0}g(t)/h(t) = 1$ with $t_0$ being the corresponding limiting point that is usually $0$ or $\oo$. The leading parameter $\kappa$ is referred to as the \emph{lower tail order} of copula $C$. In parallel, the upper tail order of copula $C$ can be defined as the $\kappa$ that satisfies that $\Chat(u,\dots,u) \sim u^\kappa \ell(u), u \ra 0^+$. Clearly, $\kappa=1$ corresponds to the usual tail dependence and then the limit of the slowly varying function $\ell(u)$ is used as the quantity to capture the degree of tail dependence.

If copulas $C_1$ and $C_2$ have lower tail orders $\kappa_1$ and $\kappa_2$ respectively and
$\kappa_1<\kappa_2$, then $C_1(\uu)\ge C_2(\uu)$ for all $\uu$ in a 
neighborhood of $\bm{0}$. Similarly, this holds for the upper tail orders
in terms of survival functions. That is, the tail order for comparing copulas implies a form of multivariate stochastic order in the joint lower or
upper tails.

To the best of our knowledge, \cite{Ledford1996} is the first paper that employs a regularly varying function to study the weaker dependence in the tails with the tail dependence parameter $\lambda = 0$. More specifically, for a bivariate random vector $(X_1 , X_2)^{\T}$, where $X_1$ and $X_2$ are unit Fr\'{e}chet distributed with cdf $F_i(x)=e^{-1/x}, x \geq 0, i=1,2$, and are nonnegatively associated, assume $\P [X_1 > r , X_2 > r] \sim \ell(r) r^{-1/\eta}, r \ra \oo$, where $1/2 \leq \eta \leq 1$. It can be verified that the tail order $\tailo$ corresponds to $1/\eta$ of Ledford and Tawn's representation, and $\eta$ is called the residual dependence index in \cite{Hashorva2010a} and references therein. A lot of research has been done following this direction. We
refer to \cite{Ledford1996,Ledford1997,Coles1999,Heffernan2000,Ramos2009} for further development of this idea.

Although the concept of tail order is defined with respect to a copula, which is a more intuitive way, the $\kappa$ itself or some functional forms of $\kappa$ simply describe the relative speed of decay of the joint tail probability to certain functional forms of the tail probability of one of the standardized margins. How to standardize the margins and how to choose the functional forms depend on how to make such relative speed of decay meaningful and be able to capture the leading information of dependence in the tails.

Another notion that is close to the concept of tail order is a tail dependence measure in the sense of \cite{Coles1999}, in which an upper tail dependence measure for a bivariate copula is defined as 
\begin{align}
	\bar{\chi}:= \lim_{u \ra 0^+} 2 \log(u)/\log(\Cbar(1-u,1-u))-1.
	\notag
\end{align}
So, $\bar{\chi} = 2/\kappa_U-1$. Note that if $C(u,\dots,u) \sim u^\kappa \ell(u)$ as $u \ra 0^+$, letting $C'(u,\dots,u):=d(C(u,\dots,u))/du$ and $[u^\kappa \ell(u)]':= d (u^\kappa \ell(u))/du$, then by the l'Hopital's rule and the Monotone density theorem \citep{Bingham1987},
\begin{eqnarray}
\lim_{u\ra0^+} \frac{\log(C(u,\dots,u))}{\log(u)} &=& \lim_{u \ra 0^+} \frac{C'(u,\dots,u)}{[u^\kappa \ell(u)]'} \times \frac{u [u^\kappa \ell(u)]'}{C(u,\dots,u)}\nonumber\\
 &=& \lim_{u\ra0^+} \frac{\kappa u^\kappa \ell(u)}{C(u,\dots, u)} = \kappa. \label{eq-alterTailorder}
\end{eqnarray}	
Therefore, in some cases it may be easier to obtain the lower tail order by applying (\ref{eq-alterTailorder}); for example, it will be used in the proof of Proposition \ref{prop-tailoGumbelMDAell} for a bivariate elliptical copula. Similarly, the upper tail order can be calculated as $\kappa_U=\lim_{u\ra 0^+}[\log(\Cbar(1-u,\dots,1-u))]/[\log(u)]$. 

In what follows, we will introduce the concepts of tail order and intermediate tail dependence in Section \ref{sect-pre}. Some detailed results for extreme value, elliptical and Archimedean copulas are presented in Sections \ref{sect-extr}, \ref{sect-elli} and \ref{sect-arch}, respectively. In particular, in Section \ref{sect-arch}, we will study intermediate tail dependence through two different stochastic representations of Archimedean copulas: the frailty model and the scale mixture model. The study of upper intermediate tail dependence of Archimedean copulas derived from the scale mixture model is new in the literature. References are given for existing results, and proofs are only provided for new results. Section \ref{sect-final} will conclude the article.

\section{Tail order and intermediate tail dependence}\label{sect-pre}
The theory of regular variation will be applied throughout the article. We refer the reader to \cite{Bingham1987}, \cite{Resnick1987,Resnick2007}, \cite{Haan2006a} and \cite{Geluk1987} for references. A measurable function $g:\Re_+ \ra \Re_+$ is regularly varying at
$\infty$ with index $\a \neq 0$ (written $g \in \RV_{\a}$) if for any $t > 0$, $\lim_{x \ra \infty} (g(xt)/g(x)) = t^{\a}$. If the above equation holds with $\a = 0$ for any $t >0$, then $g$ is said to be slowly varying at $\infty$ and written as $g \in \RV_0$. For regularly varying at $0$, that is, $\lim_{x \ra 0^+} (g(xt)/g(x)) = t^{\a}$ for any $t>0$, the notation is $g \in \RV_\al(0^+)$, and slow variation of $\ell$ at $0$ is written as $\ell \in \RV_0(0^+)$. For any $g \in \RV_{\al}, \al \in \Re$, there exists an $\ell \in \RV_0$ such that $g(x)=x^\al \ell(x)$.

Other notation: a bold letter is used to represent a transposed vector, e.g., $\xx:=(x_1,\dots,x_d)$, and $(u\one_d)$ represents $(u,\dots,u)$ with $d$ components of $u$'s and $(u\wbf)$ represents $(uw_1,\dots,uw_d)$. We use $(\wbf_I):=(w_i; i \in I)$ when $I$ is a subset of $\{1,\ldots,d\}$, and sometimes $(w_1,\dots,w_d)$ is abbreviated as $(\wbf_d)$.

\begin{definition} \label{df-tailorder} \rm
Suppose $C$ is a $d$-dimensional copula. 
If there exists a real constant $\tailo_L (C) > 0$ and $\ell \in \RV_0(0^+)$ such that 
\begin{align}
	C(u\one_d) \sim u^{\tailorder_L (C)} \ell(u), \quad u \ra 0^+, \notag  
\end{align}
then we refer to $\tailorder_L (C)$ as the \emph{lower tail order} of $C$ and refer to
$\lm_L(C) = \lim_{u \ra 0^+} \ell(u)$ as the \emph{lower tail order parameter}, provided that the limit exists. Similarly, the \emph{upper tail order} is defined as $\tailorder_{U} (C)$ that satisfies $\Cbar((1 - u)\one_d) \sim u^{\tailorder_U (C)} \ell(u), u \ra 0^+$, with the \emph{upper tail order parameter} being $\lm_U(C) = \lim_{u \ra 0^+} \ell(u)$, provided that the limit exists.
\end{definition}

The notion of tail order is especially useful for bivariate copulas or for multivariate permutation symmetric (i.e., exchangeable) copulas. Otherwise, the study of tail order involves more technical issues. All the copulas studied in this article are assumed to be permutation symmetric, in order to illustrate the main ideas and key results without involving too much technical discussion. Under such assumptions, intermediate tail dependence simply means that the corresponding tail order $\kappa$ satisfies $1 < \kappa < d$.

\begin{definition}\label{def-tailofun} \rm
Suppose $C$ is a $d$-dimensional copula and $C(u\one_d) \sim u^\tailorder
\ell(u)$, $u \ra 0^+$ for some $\ell \in \RV_0(0^+)$. The {\it lower tail
order function} $b: \mathbb{R}^d_{+} \rightarrow \mathbb{R}_+$ is defined as
  $$b(\wbf; C, \tailorder) = \lim_{u \ra 0^+} \frac{C(u w_j , 1 \leq j \leq d)}{u^{\tailorder} \ell(u)},$$
provided that the limit function exists. 
In parallel, if $\Cbar((1 - u)\one_d) \sim u^\tailorder \ell(u)$, $u \ra 0^+$ for some $\ell \in \RV_0(0^+)$, the {\it upper tail order function} $b^*: \mathbb{R}^d_{+} \rightarrow \mathbb{R}_+$ is defined as
 $$b^*(\wbf; C, \tailorder) = \lim_{u \ra 0^+} \frac{\Cbar(1-u w_j , 1 \leq j \leq d)}{u^{\tailorder} \ell(u)},$$
provided that the limit function exists. If $\ell(u) \ra h \neq 0$, then $hb(\wbf; C, 1)$ and $hb^*(\wbf; C, 1)$ become the tail dependence functions proposed in \cite{Joe2010b}. That is, the definition of the tail order function here absorbs a constant
into $\ell$ so that $b({\bf 1}_d;C,\kappa)=1$.
\end{definition}

The following are some elementary properties of the lower and upper tail order functions $b$ and  $b^*$.
Obvious properties of tail order for $\widehat{C}$ are the following:
$\tailo_L(C)=\tailo_U(\widehat{C})$, $\tailo_U(C)=\tailo_L(\widehat{C})$,
$b(\wbf;C,\tailo)=b^*(\wbf;\widehat{C},\tailo)$ and
$b^*(\wbf;C,\tailo)=b(\wbf;\widehat{C},\tailo)$. 

\begin{prop}\label{prop-basic}
A lower tail order function $b(\wbf)=b(\wbf; C, \tailorder)$ has the following properties: (1) $b(\wbf) = 0$ if there exists an $i \in \{1,\ldots, d\}$ with $w_i = 0$; (2) $b(\wbf)$ is increasing in $w_i, i \in \{1, \ldots, d\}$; (3) $b(\wbf)$ is positive homogeneous of order $\kappa$; that is, for any fixed $t > 0$, $b(t\wbf) = t^\tailorder b(\wbf)$; (4) if $b(\wbf)$ is partially differentiable with respect to each $w_i$ on $(0, + \oo)$, then $b(\wbf) = \frac{1}{\tailorder} \sum_{j=1}^d \frac{\partial
b}{\partial w_j} w_j, \forall \wbf \in \mathbb{R}^d_+.$
\end{prop}

Other properties of tail order and tail order functions can be found in \cite{Hua2011b}. Due to the limitation of space, in what follows, we will focus on the study of intermediate tail dependence for three important copula families.

\section{Extreme value copula}\label{sect-extr}
If a copula $C$ satisfies $C(u_1^t , \dots , u_d^t)=C^t(u_1, \dots, u_d )$
for any $(u_1, \dots, u_d) \in [0,1]^d$ and $t > 0$, then we refer to $C$ as an extreme value copula. For any extreme value copula $C$, there exists a function $A: [0,\oo)^d \ra [0,\oo)$ such that
\begin{align}
	C(u_1 , \dots , u_d) = \exp\{ -A(-\log u_1 , \dots , -\log u_d) \}, \label{eq-extreAfunction}
\end{align}
where $A$ is convex, positive homogeneous of order $1$ and satisfies $$\max(x_1, \dots, x_d) \leq A(x_1,\dots,x_d) \leq x_1+\dots+x_d.$$ We refer to \cite{Pickands1981}, \cite{Tawn1990} and Chapter 6 of \cite{Joe1997} for references.

For the lower tail of an extreme value copula $C$, it can be verified \citep{Hua2011b} that $C(u, \dots, u) = u^{A(1,\dots,1)}$. That is, for any extreme value copula $C$, the lower tail order is $\tailorder_L(C) =
A(1,\dots,1)$, and there is intermediate lower tail dependence except for the
boundary cases, such as the independence copula and the comonotonicity copula, where $A(1,\dots,1)=d$ and $1$, respectively.

For the upper tail of an extreme value copula $C$, write the survival copula as
$\Chat(u\one_d):=\Cbar((1-u)\one_d)= 1 + \sum_{\emptyset \neq I \subseteq \{1,\dots,d\}}(-1)^{|I|}C_I((1-u)\one_{|I|})$. Since each $I$-margin of $C$ with $2 \leq |I|\leq d$ is also an extreme value copula \citep{Tawn1990}, let $A_I$ be the corresponding function in the sense of (\ref{eq-extreAfunction}) for the extreme value copula $C_I$, then, 
\begin{align}
	\Chat(u, \dots, u) & = \Cbar(1-u, \dots, 1 - u) = 1 + \sum_{\emptyset \neq I \subseteq \{1,\dots, d\}} (-1)^{|I|} C_I(1 - \uu_{|I|}) \notag \\
	& = 1 - d + du + \sum_{I \subseteq \{1,\dots, d\}, |I|\geq 2} (-1)^{|I|} C_I(1 - \uu_{|I|}) \notag \\
	&= 1 - d + du + \sum_{I \subseteq \{1,\dots, d\}, |I|\geq 2} (-1)^{|I|} (1-u)^{A_I(\one_{|I|})}. \notag
\end{align}
By the l'Hopital's rule, as 
$u \ra 0^+$, 
$$\Chat(u,\dots,u) \sim u \left(d - \sum_{I \subseteq \{1,\dots, d\}, |I|\geq 2} (-1)^{|I|} A_I(\one_{|I|}) \right)=:u\lambda.$$
 So if $\lambda \neq 0$, then $C$ has usual upper tail dependence with tail dependence parameter $\lambda = d - \sum_{I \subseteq \{1,\dots, d\}, |I|\geq 2} (-1)^{|I|} A_I(\one_{|I|})$. If $\lambda = 0$, then it is unclear if an extreme value copula can have upper intermediate tail dependence or not; certain structures of those $A_I$ functions are needed in this regard.

\section{Elliptical copula}\label{sect-elli}
Since a copula is invariant to a strict increasing transformation on margins, for the study of elliptical copula, we may omit the location and scale parameters of joint elliptical distributions. Intermediate tail dependence depends just on the radial random variable, and the condition on the radial random variable can be seen from the bivariate case, so the main result in this section is bivariate, and it can be easily extended to exchangeable multivariate elliptical copulas. Now consider the following representation: let $\XX:=(X_1, X_2)$ be an elliptical random vector such that
\begin{align}
	\XX \eqd RA\bm{U}, \label{eq-biellrep}
\end{align}
where the radial random variable $R\geq 0$ is independent of $\bm{U}$, $\bm{U}$ is an bivariate random vector uniformly distributed on the surface of the unit hypersphere $\{\zz \in \Re^2 | \zz^{\T}\zz = 1\}$, $A$ is a $2 \times 2$ matrix such that $AA^{\T} = \Sigma$ where the entries of $\Sigma$ are $\Sigma_{11}=\Sigma_{22}=1$ and $\Sigma_{12}=\Sigma_{21}=\varrho$ with $-1<\varrho<1$, e.g., $A=\left( \begin{array}{cc}
1 & 0\\
\varrho & \sqrt{1-\varrho^2}\end{array} \right) $. For such an elliptical distribution, the margins have the same cdf assumed to be $F$.

For the usual tail dependence case, \cite{Schmidt2002} proved that when the radial random variable $R$ has a regularly varying tail then $X_1$ and $X_2$ are tail dependent, and thus the tail order of the corresponding elliptical copula is $\kappa = 1$. 

\begin{exm}\rm
(Student $t$ copula) The radial random variable $R$ for Student $t$ distributions is a generalized inverse Gamma distribution such that $R^2$ follows an inverse Gamma distribution with the shape and scale parameters being $\nu/2$, where $\nu$ is the degree of freedom. It can be verified that $\Fbar_R \in \RV_{-\nu}$ (see Example 3 of \cite{Hua2011c}). So the tail order for Student $t$ copula is $\kappa=1$.
\end{exm}

For univariate tail heaviness, one often use a concept referred to as Maximum Domain of Attraction (MDA) of a univariate extreme value distribution. The following well known result characterizes the distributions that belong to the MDA of Gumbel, which is relevant to intermediate tail dependence of elliptical copulas. We refer to \cite{Embrechts1997} and \cite{Haan2006a} for more details about MDA. For notation, $\Lambda$ is the Gumbel extreme value distribution for maxima, and $\Phi_\alpha$ is the Fr\'echet extreme value distribution
with parameter $\alpha>0$.

\begin{thm}
\label{thm-GumbelGum}
A random variable $X$ with cdf $F$ is said to belong to the Gumbel MDA (denoted as $X \in \MDA(\Lambda)$ or $F \in \MDA(\Lambda)$) if and only if there exists a positive auxiliary function $a(\cdot)$ such that
\begin{align}
	\lim_{x \ra \oo} \frac{\Fbar(x+ta(x))}{\Fbar(x)} = e^{-t}, \quad t \in \Re, \label{def-MDAGumbel}
\end{align}
where $a(\cdot)$ can be chosen as $a(x)= \int_x^\oo \Fbar(t)/\Fbar(x) dt$.
\end{thm}

For the case where $R$ has lighter tails than any regularly varying tails, some asymptotic study has been conducted for elliptical distributions where $R \in \MDA(\Lambda)$. We refer to \cite{Hashorva2007}, \cite{Hashorva2010a} and \cite{Hashorva2008} for relevant references. Now we are ready to present a result that is useful to find the tail order of a bivariate elliptical copula where the radial random variable $R$ belongs to Gumbel $\MDA$, and this result is cited from \cite{Hua2012b}. \cite{Hashorva2010a} has a version of this result in Theorem 2.1 but the proof here is different.

\begin{prop}\label{prop-tailoGumbelMDAell}
Let $C$ be the copula for an elliptical random vector $\XX:=(X_1, X_2)$ constructed as (\ref{eq-biellrep}), and $b_{\varrho} = \sqrt{2/(1+\varrho)}$. If $R \in \MDA(\Lambda)$, then the upper and lower tail orders of $C$ is
\begin{align}
	\kappa = \lim_{r \ra \oo}\frac{\log \left(1-F_R(b_{\varrho} r)\right)}{\log \left(1 - F_R(r) \right)}, \label{eq-tailoMDAGum}
\end{align}
provided that the limit exists.
\end{prop}
\pf Letting $r:= F^{-1}(1-u)$ and $b_{\varrho} = \sqrt{2/(1+\varrho)}$, then by Example 6.2 (i) of \cite{Hashorva2007}, as $u \ra 0^+$ and thus $r \ra \oo$,
\begin{align}
	&\Cbar(1-u, 1-u) \notag \\
	&= \P\left[X_1>F^{-1}(1-u), X_2>F^{-1}(1-u)\right] = \P[X_1 > r, X_2>r] \notag \\
	&= (1+o(1))\frac{(1-\varrho^2)^{3/2}}{2 \pi (1-\varrho)^2}[a(b_{\varrho}r)/r][1-F_R(b_{\varrho} r)], \label{eq-upexpasEll}
\end{align}
where $F_R$ is the cdf of $R$ and $a(\cdot)$ is an auxiliary function of $R$ with respect to the Gumbel MDA in the sense of (\ref{def-MDAGumbel}). As $u \ra 0^+$, i.e., $r \ra \oo$, both $a(b_{\varrho}r)/r \ra 0$ (see Theorems 3.3.26 and 3.3.27 of \cite{Embrechts1997}) and $1-F_R(b_{\varrho} r) \ra 0$. Let $G(x):= 1/[1-F_R(x)]$, then $G: \Re \ra \Re_+$ is increasing and the condition of (\ref{def-MDAGumbel}) is equivalent to that $G \in \Gamma$-varying with auxiliary function $a(\cdot)$ \citep[Definition 1.5.1]{Haan1970}. The inverse function of a $\Gamma$-varying function is a $\Pi$-varying function \citep[Corollary 1.10]{Haan1974}. Therefore, $G^{-1} \in \Pi$-varying. Assuming that an auxiliary function of $G^{-1}$ is $a_0(\cdot)$, by Lemma 1.2.9 of \cite{Haan2006a}, the auxiliary function $a_0(\cdot)$ of the $\Pi$-varying function $G^{-1}$ is slowly varying at $\oo$. Moreover, $a_0(t)=a(G^{-1}(t))$ \cite[Corollary 1.10]{Haan1974}. So, $a(x) = a_0(G(x))$. Then in (\ref{eq-upexpasEll}),
\begin{align}
	a(b_{\varrho} r)/r = a_0(G(b_{\varrho} r))/r = a_0(1/[1-F_R(b_{\varrho}r)])/r,\notag
\end{align}
while $1-F_R(b_{\varrho}r)$ is rapidly varying in $r$ at $\oo$ due to the fact that $G$ is $\Gamma$-varying and any $\Gamma$-varying function is rapidly varying \citep[Theorem 1.5.1]{Haan1970}. Therefore, $$1-F_R(b_{\varrho}r) = 1 - F_R\left(\sqrt{2/(1+\varrho)} F^{-1}(1-u) \right)$$ 
dominates the tail behavior of (\ref{eq-upexpasEll}) as $u \ra 0$, and thus determines the corresponding tail order of the elliptical copula. By the definition of tail order in Definition \ref{df-tailorder}, we may also obtain the upper tail order by the following
\begin{align}
	\kappa = \lim_{u \ra 0^+}\frac{\log \Cbar(1-u, 1-u)}{ \log u} = \lim_{r \ra \oo} \frac{\log \left(1-F_R(b_{\varrho} r)\right)}{\log \left(1 - F(r) \right)}. \notag
\end{align}
By Example 6.2 (iii) of \cite{Hashorva2007}, as $r \ra \oo$,
\begin{align}
	\P[X_1 > r] = (1+o(1)) (2\pi)^{-1/2} [a(r)/r]^{1/2} [1-F_R(r)]. \notag
\end{align}
Due to the similar argument as before, $1-F_R(r)$ dominates the tail behavior of $\P[X_1>r]$, as $r \ra \oo$. Therefore, we may write
\begin{align}
	\kappa = \lim_{r \ra \oo}\frac{\log \left(1-F_R(b_{\varrho} r)\right)}{\log \left(1 - F_R(r) \right)},\notag
\end{align}
which completes the proof. \hfill $\square$

It is very convenient to apply this method to derive the tail order if we know the tail behavior of $R$, and $R$ belongs to MDA of Gumbel. By Theorem 3.1 of \cite{Hashorva2007}, this result can also be extended to multivariate cases. For $d$-dimensional exchangeable elliptical copula, of which the off-diagonal entries of $\Sigma$ are all $\varrho$ and the diagonals are all $1$'s, the tail order is (\ref{eq-tailoMDAGum}) where $b_\varrho$ is replaced by $b_{\varrho, d} = \sqrt{d/[1+(d-1)\varrho]}$.

\begin{exm} \rm
(Bivariate symmetric Kotz type \citep{Fang1990} copula) The density generator 
$$g(x)=K x^{N-1} \exp\{-\beta x^\xi\}, \quad \beta,\xi,N >0,$$
where $K$ is a normalizing constant. By Theorem 2.9 of \cite{Fang1990}, the density function of $R$ is $f_R(x) = 2 \pi x g(x^2) = 2K \pi x^{2N-1} \exp\{ -\beta x^{2\xi} \}$. So, the survival function is
\begin{align}
	1-F_R(x) &= \int_x^\oo 2K \pi t^{2N-1} \exp\{ -\beta t^{2\xi} \} dt \notag \\
	& = \int_{\beta x^{2\xi}}^\oo \frac{K\pi}{\xi}\beta^{-N/\xi} w^{N/\xi-1} \exp\{-w\} dw \notag \\
	& = \frac{K\pi}{\xi}\beta^{-N/\xi} \Gamma(N/\xi, \beta x^{2\xi}), \quad \Gamma(\cdot, \cdot) \mbox{ incomplete Gamma function} \notag \\
	& \sim \frac{K\pi}{\xi}\beta^{-1} x^{2N-2\xi} \exp\{-\beta x^{2\xi}\}, \quad x \ra \oo, \notag
\end{align}
where the asymptotic relation is referred to Section 6.5 of \cite{Abramowitz1964}. Then by (\ref{eq-tailoMDAGum}), we can easily get that
\begin{align}
	\kappa = b_{\varrho}^{2\xi} = [2/(1+\varrho)]^\xi. \notag
\end{align}
Therefore, the tail order for the symmetric Kotz type copula is $\kappa = [2/(1+\varrho)]^\xi$. Gaussian copula belongs to this class with $\xi=1$, so its tail order is $2/(1+\varrho)$ which is consistent to Example 1 of \cite{Hua2011b}.
\end{exm}

\section{Archimedean copula}\label{sect-arch}
An Archimedean copula $C$ has the following typical form
\begin{align}
	C(u_1, \dots, u_d) = \psi(\psi^{-1}(u_1), \cdots, \psi^{-1}(u_d)), \label{eq-ArchCopula}
\end{align}
where $\psi$ is referred to as the generator of the Archimedean copula $C$, and $\psi$ needs to satisfy certain conditions (some papers or books also refer $\varphi = \psi^{-1}$ as the generator; e.g., \cite{Nelsen2006}). Most of the commonly used Archimedean copulas correspond to $\psi$ being a Laplace Transform (LT) of a positive random variable $X$ \citep{Joe1997, Joe1996}; that is, $\psi(s):= \int_0^\oo \exp\{ -st\}F_X(dt)$, where $F_X$ is the cdf of $X$. In this case, $\psi$ is completely monotonic, $\psi(x)$ is decreasing in $x$, $\psi(0)=1$ and $\psi(\oo)=0$. When $\psi$ is of such a LT, it can generate Archimedean copulas of any dimension. However, this condition is not necessary for a $d$-dimensional Archimedean copula with $d$ being given and finite. A necessary and sufficient condition for finite dimensional Archimedean copulas has been given in \cite{Malov2001} and \cite{McNeil2009}. 

The above two sets of conditions on $\psi$ correspond to two types of stochastic representations of Archimedean copulas, respectively. One is the well-known frailty model \citep{Marshall1988,Oakes1989}, and the other is the recently studied scale mixture model \citep{McNeil2009} (or the resource sharing model in the sense of \cite{Genest2011}) for finite dimensional Archimedean copulas. In this chapter, we will discuss conditions that lead to intermediate tail dependence of Archimedean copulas for both stochastic representations. The former has been studied in the literature \citep{Hua2011b}, and the latter is only studied for the lower tail \citep{Larsson2011}. In what follows, we will first review existing studies on intermediate tail dependence for Archimedean copulas, then we will present our findings of conditions that lead to upper intermediate tail dependence of Archimedean copulas through the scale mixture model.

\subsection{Resilience or frailty models}
Write $G_j(u):=\exp\{-\psi^{-1}(u)\}$ ($0\le u\le 1$) for $j=1,\dots,d$, then clearly $G_j$'s are identical univariate cdfs. Then, 
\begin{align}
C(u_1,\dots,u_d)=\int_0^\oo \prod_{j=1}^d G_j^\xi(u_j) \,F_H(d\xi), \label{eq-resiliArch}
\end{align}
where $F_H$ is the cdf of the resilience random variable $H$, and $\psi(s)=\psi_H(s)=\int_0^\oo e^{-s\xi} F_H(d\xi)$. The mixture representation means that: there are random variables $X_1,\ldots,X_d$ such that given $H=\xi$, they are conditionally independent with respective cdfs $G_1^\xi,\ldots,G_d^\xi$. Let $F_j(1-u):= \Gbar_j(u)$, then $F_j$'s are also cdfs, and $C(u_1,\dots,u_d)=\int_0^\oo \prod_{j=1}^d \Fbar_j^\xi(1-u_j) \,F_H(d\xi)$. Assume that $F_j$ is the cdf of $Y_j$ for each $j$. Then we can also look at the copula $C$ as the survival copula for the vector $(Y_1, \dots, Y_d)$ that are independent conditioning on the frailty random variable $H$; that is why we refer to this representation as a frailty model \citep{Genest2011}. Actually, from different perspectives, the same random variable $H$ can be referred to as either a resilience or a frailty random variable. We refer to \cite{Marshall2007} for more discussion about the concepts of resilience and frailty.

We now use the resilience model (\ref{eq-resiliArch}) to study how tail heaviness of $H$ affects the tail behavior of the corresponding Archimedean copula, and we refer to \cite{Hua2011b} for relevant studies. For any given $(u_1, \dots, u_d), 0 \leq u_i \leq 1$, a larger value of $\xi$ leads to a smaller value of $C(u_1, \dots, u_d)$, and thus a heavier right tail of $H$ tends to generate stronger positive dependence in the upper tail. Sufficient conditions on the tail heaviness of $H$ have been given in order to get an intermediate upper tail dependent Archimedean copula. If we use $M_H: = \sup\{ m \geq 0: \E[H^m] < \oo \}$ to describe the degree of tail heaviness of $H$, then under certain regularity conditions, $k < M_H < k+1$ with $k \in \{1,\dots, d-1\}$ leads to upper intermediate tail dependence. The next result is presented in \cite{Hua2011b}.

\begin{prop} \label{prop-upperArchResili}
Suppose $\psi$ is the LT of a positive random variable $Y$ with $k < M_Y <k+1$ for some $k \in \{ 1, \dots , d-1 \}$, and $|\psi^{(k)}(0) -
\psi^{(k)}(\cdot)|$ is regularly varying at $0^+$ with the associated slowly varying function $\ell$ satisfying $\lim_{s \ra 0^+}\ell(s) < \oo$. Then the Archimedean copula $C_\psi$ has upper intermediate tail dependence, and the corresponding upper tail order is $\tailorder_U =M_Y$.
\end{prop}

There are some upper intermediate tail dependent Archimedean copulas that have a simple form. One example is the Archimedean copula constructed by LT \citep{Joe2000} $$\psi(s)= \int_s^\oo \exp\{- v^\alpha\}dv / \Gamma(1+\alpha^{-1}), \quad 0 < \alpha < 1,$$
and the upper tail order is $\kappa_U=1+\alpha$ and the lower tail order is $\kappa_L=d^\alpha$. Another one-parameter Archimedean copula that has a very flexible upper tail dependence structure is the following

\begin{example} \label{ex-invgamLT} \rm
(Archimedean copula based on inverse Gamma LT (ACIG) \cite{Hua2011b}) Let $Y=X^{-1}$ and $X$ follows ${\rm Gamma}(\a,1)$ with $\a>0$, and then $M_Y=\a$ and the LT of the inverse Gamma distributed $Y$ is
\begin{align}
	\psi (s;\a) = \frac{2}{\G(\a)} s^{\a / 2} K_{\a} (2
\sqrt{s}), \quad s\ge 0,\  \a>0, \label{LTIGamma}
\end{align}
where $K_\a$ is the modified Bessel function of the second kind. For $\a\in (0, +\oo)$ noninteger, $\kappa_U(C_\psi) = \max\{1,\min\{\a,d\}\}$ and $\kappa_L(C_\psi) = \sqrt{d}$.
\end{example}

The interesting property of ACIG is that, it captures a very wide range of upper tail dependence patterns by involving only one parameter. That is, when $0<\alpha \leq 1$, there is usual tail dependence in the upper tail; when $1 < \alpha < d$, intermediate upper tail dependence is present; when $d \leq \alpha$, the upper tail becomes tail quadrant independent (i.e., $\kappa=d$).

The condition of $M_Y$ in Proposition \ref{prop-upperArchResili} being noninteger seems to be unnecessary. In the next subsection, the restriction will be relaxed when we study the tail behavior of Archimedean copulas through the scale mixture model. 

\subsection{Scale mixture models}
An Archimedean copula can also be represented as the survival copula for a random vector \citep{McNeil2009} 
\begin{align}
	\XX:=(X_1,\dots,X_d) \eqd R \times (S_1, \dots, S_d), \label{eq-Rtimessimplex}
\end{align}
where $R$ and $S_i$ are independent for $i=1,\dots,d$, $R$ is a positive random variable and $(S_1,\dots,S_d)$ is uniformly distributed on the simplex $\{\xx \in \Re^d_+: \sum_i x_i =1 \}$. The relationship between $R$ in (\ref{eq-Rtimessimplex}) and $H$ in (\ref{eq-resiliArch}) is given in Proposition 1 of \cite{McNeil2010}; that is,
\begin{align}
	R \eqd E_d /H, \notag
\end{align}
where $E_d$ is independent of $H$ and is Erlang($d$) distributed (i.e., $E_d$ follows Gamma($d,1$)). The Archimedean copula can be constructed as
\begin{align}
	C_{\psi,d}(u_1, \dots, u_d):= \psi(\psi^{-1}(u_1)+\dots+\psi^{-1}(u_d)), \label{def-ArchimedeanCopula}
\end{align}
where the generator $\psi$ is the Williamson $d$-transform of cdf $F_R$ with $F_R(0)=0$ \citep{Williamson1956}; that is,
\begin{align}
	\psi(s) = \int_s^\oo (1 - s/r)^{d-1} F(dr), \quad s \in [0,\oo). \notag
\end{align}
Note that Williamson $d$-transform of a positive random variable can also lead to a generator $\psi$ that is completely monotonic (see Example \ref{exm-acig}). Throughout this subsection, assume $\varphi(x):=\psi^{-1}(x)$. If the joint cdf of $(U_1, \dots, U_d)$ is $C_{\psi,d}$, then
\begin{align}
	(X_1, \dots, X_d) \eqd (\varphi(U_1), \dots, \varphi(U_d)) \eqd R_{\psi,d} \times (S_1, \dots, S_d); \notag
\end{align}
that is, for each margin $X_i$, $\P[X_i > x] = \psi(x)$. From the proof of Theorem 1 in \cite{Larsson2011}, we know that $X_i \eqd RY$ for $i=1,\dots,d$, where $Y \sim \mbox{Kumaraswamy}(1,d-1)$; that is, $F_Y(x)=1 - (1-x)^{d-1}, x \in [0,1]$. We refer to \cite{Song1997} for more discussion of margins of a $L_p$-norm uniform distribution.

Recently, \cite{Larsson2011} has studied the tail behavior of Archimedean copulas via the scale mixture representation. In Section 5 of \cite{Larsson2011}, some tail dependence patterns have been derived except for the intermediate upper tail dependence case. We will fill the gap in this subsection.

The lower intermediate tail dependence of bivariate Archimedean copulas has been studied in Proposition 7 of \cite{Larsson2011}, and the conditions needed on $\psi$ is essentially the same as Theorem 3.3 of \cite{Charpentier2009}. A more intuitive and fairly general pattern of $\psi$ has been considered in \cite{Hua2011b}; that is, as $s \ra \oo$, $\psi(s)\sim T(s) = a_1s^q \exp\{-a_2s^{1-\beta}\}$ and $\psi'(s) \sim T'(s)$. In this case, if $0<\beta<1$, then $C_\psi$ has lower intermediate tail dependence with
$1<\tailo_L(C_\psi)=d^{1-\beta}<d$.

With survival copula for (\ref{eq-Rtimessimplex}), the upper tail of $R$ may influence the lower tail of the corresponding Archimedean copula. Using the notation of lower tail order $\kappa$ and the relation (\ref{eq-alterTailorder}), the result about the lower intermediate tail dependence in \cite{Larsson2011} becomes the following
\begin{prop}
Let $C$ be a $d$-dimensional Archimedean copula constructed as the survival function of the random vector in (\ref{eq-Rtimessimplex}) and the associated $\psi$ in (\ref{def-ArchimedeanCopula}) is differentiable. Assume further that the radial part $R$ in (\ref{eq-Rtimessimplex}) satisfies $R \in \MDA(\Lambda)$ with the
auxiliary function $a(\cdot)$ of $R$ satisfying $a \in \RV_\beta$ for some $0<\beta < 1$, and (\ref{eq-alterTailorder}) holds, then $\tailo_L = d^{1-\beta}$ and thus $C$ has lower intermediate tail dependence. 
\end{prop}

Based on (\ref{eq-Rtimessimplex}), the lower tail of $R$, or equivalently, the upper tail of $1/R$ may affect the upper tail of the associated Archimedean copula. In what follows, we will prove that $1/R$ belonging to the MDA of $\FRE$ (written as $1/R \in \MDA(\Phi_\alpha)$) may lead to upper intermediate tail dependence for the associated Archimedean copula; here the condition of the tail order being an integer is relaxed. 

\begin{prop}
\label{prop-williamsontransupperinterm}
If $1/R \in \MDA(\Phi_{\alpha})$ with $k \leq \al < k+1$, where $k \in \{1,\dots,d-1\}$ is a positive integer, and $\psi$ is the Williamson $d$-transform of $F_R$ such that $F_R(0)=0$, and if $k=d-2$ or $d-1$, then further require that $\psi$ is $(d+1)$-monotone and $(d+2)$-monotone respectively, then the Archimedean copula constructed by $\psi$ as (\ref{def-ArchimedeanCopula}) has upper tail order $\kappa_U=\al$ and thus upper intermediate tail dependence if $1 < \alpha <d$. 
\end{prop}
\pf First note that either $\psi$ being $(d+1)$-monotone or $(d+2)$-monotone can imply $\psi$ being $d$-monotone \citep{McNeil2009}. So we can still apply the Willianmson $d$-transform for the two cases where $k=d-2$ or $d-1$. By definition, the Williamson $d$-transform of $R$ is
\begin{align}
	\psi(s) &= \int_s^\oo \lss 1-s/r \rss^{d-1}F_R(dr) = \int_s^\oo \sum_{i=0}^{d-1} \binom{d-1}{i}(-s/r)^i F_R(dr) \notag \\
	        & = 1 - F_R(s) +  \sum_{i=1}^{d-1}\binom{d-1}{i}(-1)^i s^i \int_s^\oo r^{-i} F_R(dr) \notag \\
	        & = 1 - F_R(s) +  \sum_{i=1}^{d-1}\binom{d-1}{i}(-1)^i s^i \lss -s^{-i}F_R(s)+i\int_s^\oo F_R(r)r^{-i-1}dr \rss \notag \\
	        & = 1 - F_R(s) +  \sum_{i=1}^{d-1}\binom{d-1}{i}(-1)^i \lss -F_R(s)+i s^i \int_0^{1/s}F_R(1/y)y^{i-1}dy \rss \notag \\
	        &= 1 +  \sum_{i=1}^{d-1}\binom{d-1}{i}(-1)^i \lss i s^i \int_0^{1/s}F_R(1/y)y^{i-1}dy \rss \notag \\
	        &=: 1 +  \sum_{i=1}^{d-1}\binom{d-1}{i}(-1)^i m_i(s). \label{eq-Psiexpansion}
\end{align}

For any $1 \leq \alpha<d$, there exists a positive integer $k\in \{1,2, \dots,d-1\}$ such that $$k \leq \alpha < k+1.$$ 
To study the upper tail for an Archimedean copula, we now investigate the behavior of the functions $m_i(s)$ in (\ref{eq-Psiexpansion}) for $i=k,\dots,d-1$, as $s \ra 0^+$. Depending on the value of $i$, we consider the following cases: 

\textbf{Case 1:} $\alpha < k+1 \leq i \leq d-1$. 

The condition $1/R \in \MDA(\Phi_\al)$ implies that $\P[1/R > \cdot] \in \RV_{-\al}$, and $F_R = \P[R \leq \cdot] \in \RV_{\al}(0^+)$. Write $F_R(s):= s^\al \ell_{R}(s)$, where $\ell_{R} \in \RV_0(0^+)$. Since $y \mapsto F_R(1/y) \in \RV_{-\al}$, we have $y \mapsto F_R(1/y)y^{i-1} \in \RV_{i-\al-1}$. By Karamata's Theorem (e.g., \cite{Resnick2007}), $i \geq k+1 > \al$ implies that
\begin{align}
	\int_0^{1/s}F_R(1/y)y^{i-1}dy \sim \frac{1}{i-\al}F_R(s)s^{-i}, \quad s \ra 0^+, \notag 
\end{align}
and thus
\begin{align}
	m_i(s):= i s^i \int_0^{1/s}F_R(1/y)y^{i-1}dy \sim \frac{i}{i-\al}F_R(s) = \frac{i}{i-\al} s^\al \ell_{R}(s), \quad s \ra 0^+. \label{eq-tailbehaviorofthemfunction}
\end{align}

\medskip

\textbf{Case 2:} $1 \leq i < \alpha$. The condition $F_R \in \RV_\alpha(0^+)$ implies that
\begin{align}
	\ell_i(s):=i \int_0^{1/s} F_R(1/y)y^{i-1}dy \nearrow \E[R^{-i}] < \oo, \quad \mbox{as } s \ra 0^+. \notag
\end{align}
Therefore, $m_i(s) \sim \E[R^{-i}]s^i$ as $s \ra 0^+$.

\medskip

\textbf{Case 3:} $i = k = \alpha$. Let $\ell_k(s):= k\int_0^{1/s} F_R(1/y)y^{k-1}dy$. Similar to the derivation of (\ref{eq-tailbehaviorofthemfunction}), by Karamata's theorem (e.g., Theorem 2.1 (a) of \cite{Resnick2007}), $\ell_k \in \RV_0(0^+)$, and hence, $m_k \in \RV_{k}(0^+)$.

\bigskip

Then $\psi(s) = 
1 +  \sum_{i=1}^{d-1}\binom{d-1}{i}(-1)^i m_i(s)$ implies that:

if $\alpha$ is a positive non-integer, that is, $k<\alpha$, then $\ell_i(s) \ra \E[R^{-i}]$ for $i=1, \dots, k$, and
\begin{align}
	\psi(s) \sim 1 & + \sum_{i=1}^{k}\binom{d-1}{i}(-1)^i s^i \ell_i(s)  \notag \\
	&+ \sum_{i=k+1}^{d-1}\binom{d-1}{i}(-1)^i \frac{i}{i-\alpha} s^\al \ell_{R}(s), \quad s \ra 0^+;\label{sim-noninteger}
\end{align}

if $\alpha$ is a positive integer, that is, $k=\alpha$, then $\ell_i(s) \ra \E[R^{-i}]$ for $i=1, \dots, k-1$, and
\begin{align}
	\psi(s) \sim 1 & + \one\{k>1\} \times \sum_{i=1}^{(k-1)\vee 1}\binom{d-1}{i}(-1)^i s^i \ell_i(s) \notag \\
	&+ \binom{d-1}{k}(-1)^k s^k \ell_k(s) \notag \\
	&+ \sum_{i=k+1}^{d-1}\binom{d-1}{i}(-1)^i \frac{i}{i-\alpha} s^\al \ell_{R}(s), \quad s \ra 0^+. \label{sim-integer}
\end{align}

Therefore, $1 \leq \alpha$ implies that the map $1 - \psi(\cdot) \in \RV_1(0^+)$. Write $1 - \psi(s) = s \ell_0(s)$, where $\ell_0 \in \RV_0(0^+)$. 

\medskip

Assume that there exists a constant $\kappa$ and a slowly varying function $\ell_* \in \RV_0(0^+)$ such that $\Cbar(1-u, \dots, 1-u) \sim u^\kappa \ell_*(u)$ as $u \ra 0^+$. If we can prove that $\kappa = \alpha$, get the expression of $\ell_*$, and prove that such an $\ell^*$ is a slowly varying function, then the proof is finished.

Let $s:= \psi^{-1}(1-u)$, then $1 - \psi(s) = s \ell_0(s)$ implies that
\begin{align}
	1&= \lim_{u \ra 0^+}\frac{\Cbar(1-u, \dots, 1-u)}{u^\kappa \ell_*(u)} = \lim_{u \ra 0^+} \frac{\lbb 1+\sum_{j=1}^d (-1)^j \binom{d}{j}\psi(j \psi^{-1}(1-u)) \rbb }{[u^\kappa \ell_*(u)]} \notag \\
	& = \lim_{s \ra 0^+} \lbb 1+ \sum_{j=1}^d (-1)^j \binom{d}{j}\psi(j s) \rbb / [s^\kappa \ell_0^\kappa(s) \ell_*(1-\psi(s))]. \notag
\end{align}
The locally uniform convergence of $\ell_*$ at $0^+$, together with $1 - \psi(\cdot) \in \RV_1(0^+)$ implies that, for any given $t>0$,
\begin{align}
	\lim_{s \ra 0^+} \frac{\ell_*(1-\psi(ts))}{\ell_*(1-\psi(s))} = \lim_{s \ra 0^+} \frac{\ell_*(\frac{1-\psi(ts)}{1-\psi(s)}\times (1-\psi(s))}{\ell_*(1-\psi(s))} = \lim_{u \ra 0^+} \frac{\ell_*(tu)}{\ell_*(u)} = 1. \notag
\end{align}
Hence, $\ell_*(1-\psi(\cdot)) \in \RV_0(0^+)$. Let $\ell(s):= \ell_0^\kappa(s) \ell_*(1-\psi(s))$, then $\ell \in \RV_0(0^+)$, due to Proposition 1.3.6 of \cite{Bingham1987}.

Applying the Monotone Density Theorem \citep[Theorem 1.7.2b]{Bingham1987} $k-1$ times on (\ref{eq-Psiexpansion}) implies that $\psi^{(i)}(0) < \oo$ for $i=0,\dots,k-1$. Moreover, choosing $w_i \equiv 1$ for each $i$ in Lemma 2 of \cite{Hua2011b} implies that $\sum_{j=1}^d (-1)^{j} \binom{d}{j} j^i  \equiv 0$ for any positive integer $i$ that is less than $d$. Therefore, $\sum_{j=1}^d (-1)^j \binom{d}{j} j^{i}\psi^{(i)}(0) = 0$ for $i=0,\dots,k-1$. Define $[x]:= \max\{z \mbox{ integer; } z < x\}$, and for any $y > 0$, $y!:= y\times (y-1)\times \cdots \times (y - [y])$. By the l'Hopital's rule,
\begin{align}
	1&= \lim_{s \ra 0^+} \frac{\sum_{j=1}^d (-1)^j \binom{d}{j} j^{k-1} \psi^{(k-1)}(j s) }{ [\kappa! / (\kappa-k+1)!] s^{\kappa-k+1} \ell(s)}. \label{eq-startingfromkminus1}
\end{align}

If $\alpha$ is a non-integer, that is, if $k < \alpha < k+1$, then applying the Monotone Density Theorem two more times for (\ref{eq-startingfromkminus1}) with respect to (\ref{sim-noninteger}) leads to 
\begin{align}
	(-1)^{k+1}\psi^{(k+1)}(s) \sim \left| \sum_{i=k+1}^{d-1}\binom{d-1}{i}(-1)^i \frac{i}{i-\alpha} \right| [\alpha !/(\alpha - k -1)!] s^{\al-k-1} \ell_{R}(s); \label{sim-kplus1deriv}
\end{align}

\begin{eqnarray}
	1&=& \lim_{s \ra 0^+} \frac{\sum_{j=1}^d (-1)^j \binom{d}{j} j^{k} \psi^{(k)}(j s) }{ [\kappa! / (\kappa-k)!] s^{\kappa-k} \ell(s)}\nonumber\\
	& =& \lim_{s \ra 0^+} \frac{\sum_{j=1}^d (-1)^{j-k-1} \binom{d}{j} j^{k+1} [(-1)^{k+1}\psi^{(k+1)}(j s)] }{ [\kappa! / (\kappa-k-1)!] s^{\kappa-k-1} \ell(s)} \label{eq-nonintegalpha}
\end{eqnarray}
Combining (\ref{sim-kplus1deriv}) and (\ref{eq-nonintegalpha}) leads to $\kappa = \alpha$, and then 
$$\ell(s) = \sum_{j=1}^d(-1)^{j-k-1} \binom{d}{j}j^\alpha \left| \sum_{i=k+1}^{d-1}\binom{d-1}{i}(-1)^i \frac{i}{i-\alpha} \right| \ell_R(js),$$
which is a slowly varying function. Thus, the slowly varying function $\ell^*$ can be chosen accordingly, which proves the case where $\alpha$ is a positive non-integer.

If $\alpha$ is a positive integer, that is, $\alpha = k$, then applying the Monotone Density Theorem one more time for (\ref{eq-startingfromkminus1}) with respect to (\ref{sim-integer}) leads to $\kappa = k = \alpha$, and similarly the slowly varying function $\ell^*$ can be obtained accordingly.
\QED

\begin{exm}[ACIG copula] \rm
\label{exm-acig}
For the ACIG copula studied in Example 4 of \cite{Hua2011b}, $H^{-1} \sim \mbox{Gamma}(\alpha, 1)$ for $\alpha >0$. Therefore, $R$ has the same distribution as the product of two independent 
Gamma random variables with scale parameter 1 and respective shape parameters $d$ and $\alpha$. The product follows a K-distribution and the density function is
\begin{eqnarray}
	f_R(x;d,\al) &=& \frac{2}{\Gamma(d)\Gamma(\al)}x^{(\al+d)/2-1}K_{d-\al}(2\sqrt{x}), \nonumber\\ && x \in [0,\oo);
	  d \mbox{ positive integer}; \al>0, \label{eq-densityofRforACIG}
\end{eqnarray}
where $K$ is the modified Bessel function of the second kind. We refer to \cite{Jakeman1978} for the reference of density functions of a K-distribution. Because the left tail behavior of $R$ affects the upper tail dependence pattern of the corresponding Archimedean copula derived from (\ref{eq-Rtimessimplex}), we need to study the behavior of (\ref{eq-densityofRforACIG}) at $0$. If $\al$ is not an integer, then
\begin{align}
   K_{d-\al}(s) \sim \frac{1}{2}\lss\Gamma(d-\al)(s/2)^{\al-d}+\Gamma(\al-d)(s/2)^{d-\al}\rss, \quad s \ra 0^+. \notag 	
\end{align}
Therefore,
\begin{align}
	f_R(x;d,\al) \sim \frac{\Gamma(d-\al)}{\Gamma(d)\Gamma(\al)}x^{\al-1}+\frac{\Gamma(\al-d)}{\Gamma(d)\Gamma(\al)}x^{d-1}, \quad x \ra 0^+. \notag
\end{align}
If $1<\al<d$, then the term $x^{\al-1}$ dominates the tail behavior at $0$; this is upper intermediate tail dependence case. When $d-\alpha$ is a positive integer, then by \cite{Abramowitz1964},
\begin{align}
K_{d- \alpha}(s) &\sim \half \left(s/2 \right)^{\alpha-d} \sum_{k=0}^{d-\alpha-1} {(-1)^k (d-\alpha-k-1)! \over k!}(s/2)^{2k} \notag \\
 &\sim {(d-\alpha -1)!\over 2} \left(s/2 \right)^{\alpha-d} , \quad s \ra 0^+. \notag
\end{align}
Therefore,
\begin{align}
	f_R(x;d,\al) \sim {(d-\alpha-1)! \over \Gamma(d) \Gamma(\alpha) } x^{\alpha-1}= {\Gamma(d-\alpha) \over \Gamma(d) \Gamma(\alpha) } x^{\alpha-1}, \quad x \ra 0^+, \notag
\end{align}
which is the same as $\alpha$ being a non-integer. Combining these two cases, $1<\alpha<d$ implies upper intermediate tail dependence of the ACIG copula.
\end{exm}

Proposition \ref{prop-williamsontransupperinterm} is actually very useful to guide us to find a scale mixture random vector whose survival copula is an upper intermediate tail dependent Archimedean copula. For example, let $R$ follow the positive Weibull distribution; that is, $F_R(x) = 1 - \exp\{- x^\al\} \sim x^\al, x \ra 0^+$ with $1 < \al < d$. Then Proposition \ref{prop-williamsontransupperinterm} implies that $\kappa_U = \alpha$ and thus the corresponding Archimedean copula has upper intermediate tail dependence. Another example of Archimedean copula that has upper intermediate tail dependence is presented in the following.

\begin{exm}[Dagum-simplex mixture] \rm
The Dagum distribution is also referred to as an inverse Burr distribution, and it is a special case of generalized beta distribution of the second kind (e.g., \cite{Kleiber2003}). Let the cdf of the radial random variable $R$ be Dagum, then
\begin{align}
	F_R(x) = \lmm 1+(x/\sigma)^{-\al} \rmm^{-\beta}, \quad x>0, \quad \al , \beta, \sigma>0. \notag
\end{align}
We choose $\sigma=1$ for simulation as the scale parameter does not affect the associated Archimedean copula. It can be derived that $F_R \in \RV_{\al \beta}(0^+)$. By Proposition \ref{prop-williamsontransupperinterm}, if $1 < \al\beta <2$, then the copula $C(u,v):=\psi(\psi^{-1}(u)+\psi^{-1}(v))$ should have upper tail order $\kappa_U = \al\beta$. The simulated scatter plots are illustrated in Figure \ref{fig-dagumsimplex}, where the left plot is for uniform margins, and the right plot is for standard normal margins. The sample size was $2000$ for the simulations. 
\begin{figure}[htp]
\centering
\caption{Simulation of Dagum-simplex copula}
\medskip
\label{fig-dagumsimplex}
\resizebox{6.7cm}{7.4cm}{\includegraphics{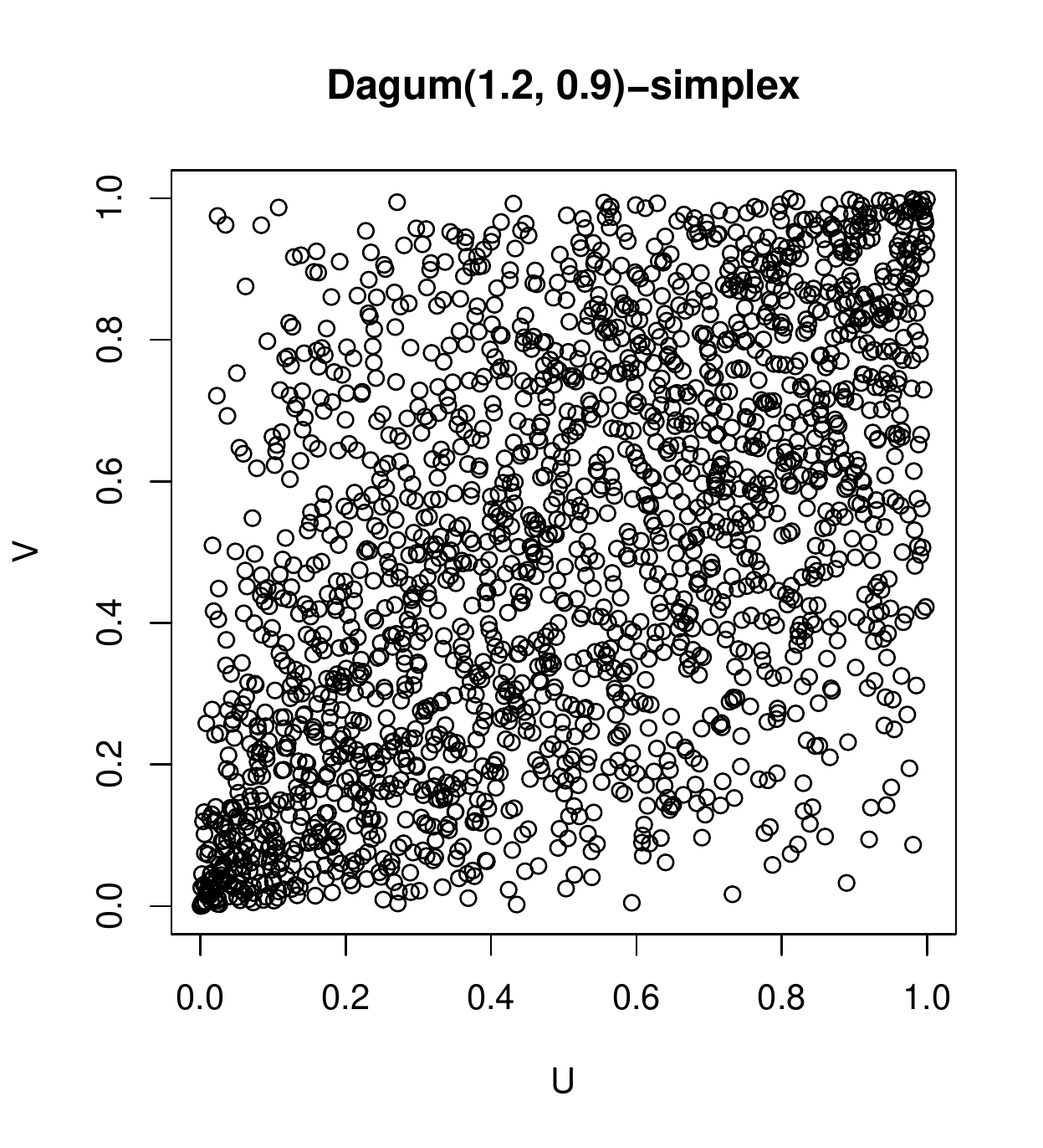}}
\resizebox{6.7cm}{7.4cm}{\includegraphics{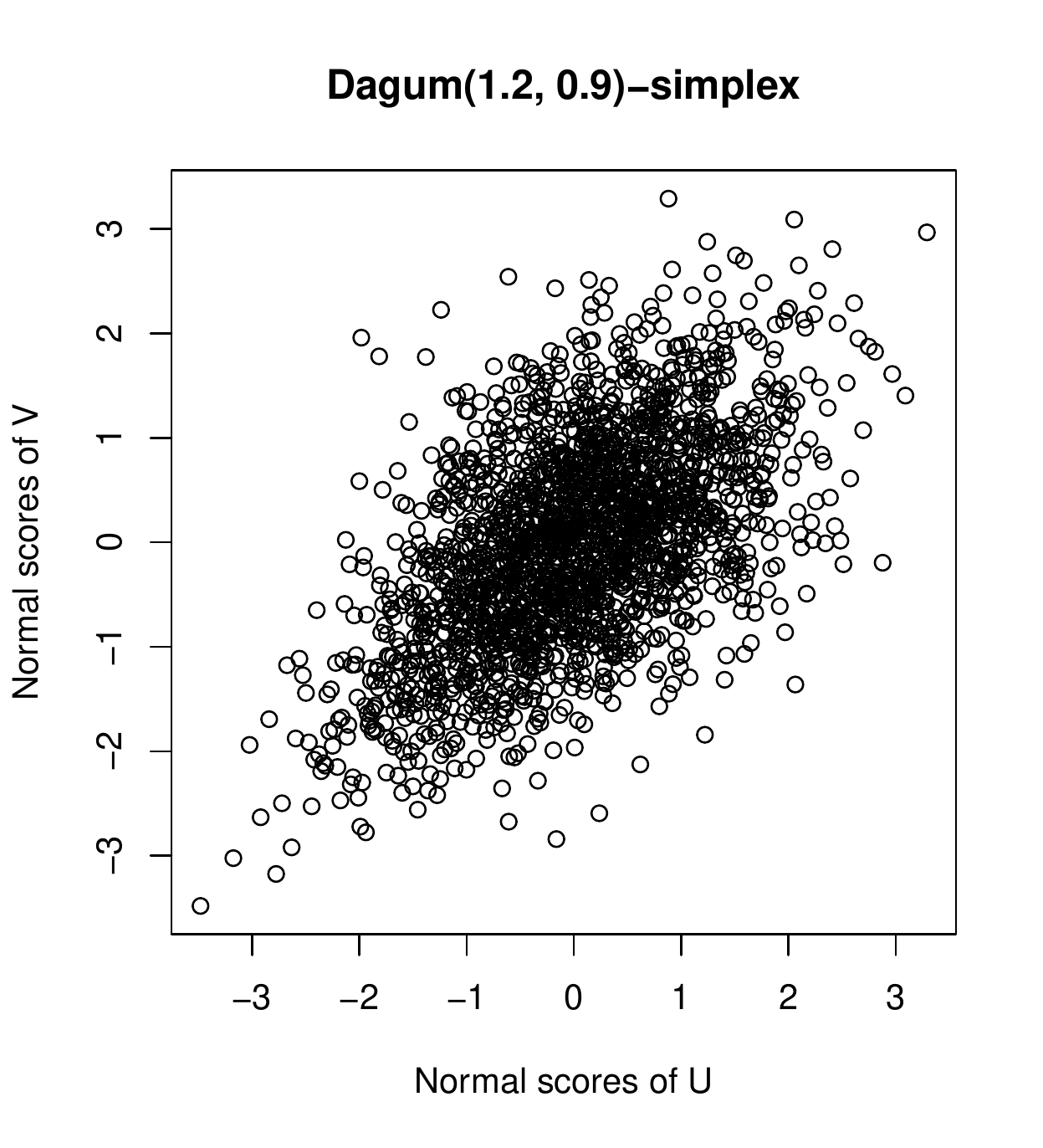}}
\end{figure}
\end{exm}
In Figure \ref{fig-dagumsimplex}, the upper tail order is $\kappa_U=1.08$, which belongs to upper intermediate tail dependence. For the lower tail, $\Fbar_R(x) = 1 - (1+x^{-\al})^{-\beta}$, and $\Fbar^{(1)}_R(x) = -\al \be (1+x^{-\al})^{-\be-1}x^{-\al-1} \sim -\al \be x^{-\al-1}$ as $x \ra \oo$. So $\Fbar_R \in \RV_{-\al}$. By Corollary 2 of \cite{Larsson2011}, there is lower tail dependence and the tail order parameter $\lambda_L = 2^{-\al}$.

\section{Remark and future work}\label{sect-final}
The notion of tail order provides a quantity to evaluate the degree of dependence in the tails of joint distributions, especially when intermediate tail dependence appears. We first review fundamental concepts and existing results of intermediate tail dependence. Throughout the review, some new properties of intermediate tail dependence have been given to supplement existing results. The new results mainly consist of an easy way to derive the tail order of a bivariate intermediate tail dependent elliptical copula, and the study of intermediate upper tail dependence for Archimedean copulas constructed from a scale mixture model. 

Proposition \ref{prop-williamsontransupperinterm} is helpful for constructing an upper intermediate tail dependent Archimedean copula. However, the scale mixture approach can often only give us a simple way to simulate desired tail dependence structures, but not a simple closed-form parametric copula family. So, how to apply the scale mixture model to provide various desirable models and to make statistical inference  efficiently will be a very interesting topic for future research.

As mentioned earlier in Section \ref{sec-intro}, there is a link between tail orders and multivariate stochastic orders. Future research also includes whether certain forms of multivariate stochastic orders \citep[Chapter 6]{Shaked2007}
can be adapted to tail forms for comparing the strength of dependence
in the tails of copulas.

\bibliographystyle{apalike} 
\bibliography{ITD}

\end{document}